\documentclass[prl,twocolumn,aps,showpacs]{revtex4}
\usepackage{amsmath}
\usepackage{epsfig}

\newcommand{\jap} {J. Appl. Phys.}
\newcommand{\ijqc} {Int. J. Quant. Chem.}

\newcommand{\xc} {exchange-correlation}

\newcommand{\intsh}{\int \!}
\newcommand{\vecr} {{\mathbf r}}
\newcommand{\veck} {{\mathbf k}}
\newcommand{\vecq} {{\mathbf q}}
\newcommand{\vecG} {{\mathbf G}}

\newcommand{\Exc} {E_{\rm xc}}

\newcommand{\vxc} {v_{\rm xc}}
\newcommand{\vx} {v_{\rm x}}

\newcommand{\dvxc} {\delta \vxc}
\newcommand{\dvx} {\delta \vx}

\newcommand{\dn} {\delta n}

\newcommand{\wC} {w_C}
\newcommand{\FH} {F_{\rm H}}
\newcommand{\FHhat} {\widehat{F}_{\rm H}}
\newcommand{\Fxc} {F_{\rm xc}}
\newcommand{\Fx} {F_{\rm x}}
\newcommand{\Hx} {H_{\rm x}}

\newcommand{\chihat} {\widehat{\chi}}
\newcommand{\chitilde} {\widetilde{\chi}}

\newcommand{\owc} {\widehat{w}_C}
\newcommand{\ovxnl} {\widehat{\Sigma}_x}
\newcommand{\ovx} {\widehat{v}_x}

\newcommand{\ak} {a\veck}
\newcommand{\bk} {b\veck}
\newcommand{\bkp} {b\veck'}

\newcommand{\skq} {s\veck+\vecq}

\newcommand{\tkq} {t\veck + \vecq}
\newcommand{\tkpq} {t\veck'+\vecq}

\newcommand{\eak} {\epsilon_{\ak}}
\newcommand{\ebk} {\epsilon_{\bk}}
\newcommand{\ebkp} {\epsilon_{\bkp}}
\newcommand{\eskq} {\epsilon_{\skq}}
\newcommand{\etkq} {\epsilon_{\tkq}}
\newcommand{\etkpq} {\epsilon_{\tkpq}}

\newcommand{\wwpid} {\omega + i \delta}

\newcommand{\lak} {\langle \ak |}
\newcommand{\rak} {| \ak \rangle}

\newcommand{\lskq} {\langle \skq |}
\newcommand{\rskq} {| \skq \rangle}
\newcommand{\lbk} {\langle \bk |}
\newcommand{\rbk} {| \bk \rangle}
\newcommand{\rbkp} {| \bkp\rangle}

\newcommand{\ltkq} {\langle \tkq|}
\newcommand{\ltkpq} {\langle \tkpq|}

\newcommand{\rtkq} {| \tkq\rangle}

\newcommand{\qzero} {q \rightarrow 0}
\newcommand{\Order} {{\mathcal O}}

\begin{document}

\title{Excitonic optical spectrum of semiconductors obtained by time-dependent
density-functional theory with the exact-exchange kernel}

\author{Yong-Hoon Kim}
\altaffiliation{Present address:
    Materials and Process Simulation Center (139-74),
    California Institute of Technology,
    Pasadena, CA 91125-7400}
\author{Andreas G{\"o}rling}
\affiliation{Lehrstuhl f{\"u}r Theoretische Chemie,
    Technische Universit{\"a}t M{\"u}nchen,
    D-85748 Garching, Germany}

\date{\today}

\begin{abstract}
Applying the novel exact-exchange (EXX) Kohn-Sham method within
time-dependent density-functional theory, we obtained the optical
absorption spectrum of bulk silicon in good agreement with
experiments including excitonic features.  Analysis of the EXX
kernel shows that inclusion of the Coulomb coupling of
electron-hole pairs and the correct long-wavelength behavior in
the kernel is crucial for the proper description of excitonic
effects in semiconductors.
\end{abstract}


\maketitle


The calculation of optical spectra of solids has been
traditionally one of the most important and challenging areas of
first-principles material investigations. The interaction of
excited electrons and holes plays a crucial role in optical
excitations, however properly incorporating such effects {\it ab
initio} is theoretically and computationally very difficult. In
the past several years there has been a major advance in the
field~\cite{BSE98} based on the solution of the Bethe-Salpeter
equation (BSE)~\cite{Hanke79} for the two-particle Green's
function starting from quasiparticles obtained within the GW
approximation (GWA)~\cite{Hybertsen85}, but unfortunately the
calculation scheme is by nature complicated and demanding.

One attractive alternative is extending density-functional theory
(DFT) to the time-dependent (TD) case~\cite{TDDFT}. TDDFT is a
well-founded theory for the treatment of electronic excitations in
general and, unlike the GWA-BSE route which is presently
implemented as a post-DFT method, provides a computational scheme
based directly on Kohn-Sham (KS) one-particle states. Moreover,
once the KS one-particle states are given, the electronic linear
response properties are completely determined by the
orbital-independent Hartree kernel $\FH(\vecr,\vecr') \equiv 1 /
|\vecr - \vecr'|$~\cite{AU} and the {\em dynamic} \xc\ kernel
\begin{equation}
\label{eq:Fx}
 \Fxc(\vecr,\vecr';t-t')
    \equiv \frac{\dvxc(\vecr;t)}{\dn(\vecr';t')}.
\end{equation}
Thus, in principle, one only needs to find a good approximation to
the \xc\ kernel for practical TDDFT applications, which represents
a significant conceptual and computational simplification. In
practice, however, the conventional adiabatic local-density
approximation (LDA) and generalized-gradient approximation (GGA)
kernels are known to be inadequate for the study of optical
excitations in solids.  A primary example is their incorrect
long-wavelength behavior for insulators which has been recently
emphasized in several contexts~\cite{Aulbur96,Ghosez97,Reining02}.

Deficiencies of the LDA and GGA in fact appear already on the
level of the {\em ground-state} \xc\ energy $\Exc$ and potential
$\vxc(\vecr) \equiv \delta \Exc/\dn(\vecr)$.  With respect to
energetics, for example, the LDA and GGA inherently fail to
describe the quasi-two-dimensional electron gas~\cite{2D-egas} and
the van der Waals interactions~\cite{vdW}. For $\vxc$ and the
corresponding KS eigenvalue spectrum,  LDA and GGA $\vxc$
incorrectly decay exponentially rather than as $-1/r$ in finite
systems and accordingly their highest occupied orbital energies
are too high and unoccupied orbitals do not exhibit Rydberg
series, and for solids their band gaps are too small.  The
deficiencies of the LDA and GGA $\vxc$ in particular pose
significant practical difficulties for the study of electronic
excitations within the DFT formalism.

In this regard, recent realizations of multi-dimensional KS
exact-exchange (EXX)~\cite{Stadele97,Gorling99,Ivanov99} and
effective EXX~\cite{Bylander95,Grabo97,KLI99,DellaSala01} methods
provide an interesting opportunity. The orbital-based
self-interaction-free EXX methods not only provide realistic local
multiplicative KS exchange potentials and KS eigenvalue spectra
for molecules~\cite{Grabo97,KLI99,Gorling99,Ivanov99,DellaSala01}
but also give band gaps of semiconductors in good agreement with
experiments~\cite{Stadele97,Bylander95}. We have also recently
shown~\cite{DFT01} that the EXX spectrum at the one-particle level
without any previously applied post-DFT modification such as a
quasiparticle shift~\cite{Levine89} gives a very good description
of the absorption spectrum of semiconductors excluding excitonic
features resulting from two-particle interactions, which supports
the notion that KS eigenvalue differences represent well-defined
approximations for excitation energies~\cite{Gorling96,Umrigar98}.
However, the remaining excitonic character in the spectrum is not
properly treated by the LDA or GGA kernel, so a complete set of
DFT methods for the investigation of electronic optical
excitations in solids is still missing.

Given the importance of proper inclusion of excitonic effects in
the study of electronic optical excitations and the encouraging
performance of the EXX method for band structures of
semiconductors, we consider in this work a TDDFT scheme based on
the EXX kernel. At the formal level, in contrast to LDA and GGA
kernels, the EXX kernel is nonlocal in real space, depends
explicitly on the frequency, and has the correct long-wavelength
limit behavior~\cite{Ghosez97,FXEXX02}. Taking bulk silicon as the
representative semiconductor system, we show that the EXX kernel
indeed provides optical spectra of semiconductors in good
agreement with experiments {\em including excitonic effects} and
further analyze the origin of the excitonic peak and the validity
of locality approximations in space and time.  We additionally
make comparisons between EXX-TDEXX and GWA-BSE, which renders new
insight into the TDDFT approach in general~\cite{Gonze99}.


To obtain the optical spectrum of a solid, we computed the
frequency $\omega$-dependent macroscopic dielectric function
$\epsilon_{\rm M}(\omega)$ which can be written in terms of the
modified full linear response matrix $\chihat$ as~\cite{Hanke79}
\begin{equation}
\label{eq:df-mac}
 \epsilon_{\rm M}(\omega)
    = 1 - \lim_{\qzero} \FH(\vecq) \chihat(\vecq;\omega)
    |_{\vecG = \vecG' = 0},
\end{equation}
where $\vecq$ is the photon momentum, and $\FH(\vecq) \equiv
\FH(\vecG,\vecG',\vecq) = \delta_{\vecG,\vecG'} 4 \pi / |\vecq +
\vecG|^2$. Within TDDFT, $\chihat$ is completely determined by the
KS linear response matrix $\chi_0$ and the kernel matrix $\Fxc$
according to
\begin{equation}
\label{eq:chihat-chi0}
 \chihat(\vecq;\omega) =
    \Bigl[ 1 - \chi_0(\vecq;\omega) \{ \FHhat + \Fxc(\vecq;\omega) \}
    \Bigr]^{-1}
    \chi_0(\vecq;\omega),
\end{equation}
where $\FHhat = 0$ for $\vecG = \vecG' = 0$ and $\FHhat = \FH$
otherwise.  At the independent-particle level, $\chihat = \chi_0$,
the two-particle interaction effects due to $\FHhat$ and $\Fxc$
are both ignored, while only the $\Fxc$ part is ignored at the
time-dependent Hartree level. As is apparent from Eqs.
(\ref{eq:df-mac}) and (\ref{eq:chihat-chi0}), the long-wavelength
$\qzero$ behavior of the ``head'' ($\vecG = \vecG' = 0$) and
``wing'' ($\vecG =0$ and $\vecG' \neq 0$ or vice versa) elements
of $\Fxc$ (in insulators, $\Order(1/q^2)$ for the head and
$\Order(1/q)$ for the wings~\cite{Ghosez97,FXEXX02}) plays an
important role.

The computation of $\chi_0$ is straightforwardly done with KS
one-particle states and energies, and consequently $\Fxc$ remains
as the only important quantity to be determined. So far, for
$\Fxc$, the adiabatic LDA kernel $\Fxc^{LDA} = \delta^2 \Exc^{LDA}
/ (\delta n \, \delta n')$ has been almost exclusively employed.
It is local in real space and frequency-independent, which results
in a reciprocal-space representation independent of $q$ and
$\omega$, $\Fxc^{LDA}(\vecG,\vecG',\vecq; \omega) =
\Fxc^{LDA}(\vecG-\vecG')$. While numerical advantageous, this
drastic simplification is known to be problematic for the study of
optical responses in solids. For example, the head and wings of
the LDA kernel are incorrectly nondivergent in the $\qzero$ limit
and this defect cannot be corrected by a semilocal GGA kernel.

Instead, we adopted in this work the EXX kernel,
\begin{equation}
\label{eq:Fx-EXX}
\begin{split}
\Fx^{EXX} & (\vecG,\vecG',\vecq;\omega) = \sum_{\vecG_1,\vecG_2}
    \chi_0^{-1}(\vecG,\vecG_1,\vecq; \omega) \\
    & \times \Hx(\vecG_1,\vecG_2,\vecq;\omega)
    \, \chi_0^{-1}(\vecG_2,\vecG',\vecq;\omega),
\end{split}
\end{equation}
which is nonlocal in real space and explicitly depends on the
frequency.  The expression of the EXX kernel ``core'' $\Hx$, which
was interpreted in the many-body diagrammatic picture as the
first-order self-energy and vertex corrections to the irreducible
polarizability $\chitilde = \chi_0 + \chi_0 \, \Fxc \,
\chitilde$~\cite{Tokatly01}, has been presented in
Ref.~\cite{FXEXX02} for the case of periodic insulators. We have
further shown that $\Fx^{EXX}$ has the $\qzero$ behavior of the
exact $\Fxc$ and thus rectifies one serious deficiency of the LDA
and GGA kernels. The full expression for $\Hx$ is rather lengthy
and we list here only the three resonant terms~\cite{convention}:
\begin{widetext}
\begin{equation}
 \label{eq:Hx-resonant}
\begin{split}
 \Hx^{A-\textrm{res}}(\vecq;\omega) \equiv
 & - \frac{2}{\Omega} \sum_{as\veck}\sum_{bt\veck'}  \Bigl[
    \frac{\lak e^{-i(\vecq+\vecG) \cdot \vecr} \rskq \
        \langle \skq; \bkp |\owc| \tkpq; \ak \rangle
    \  \ltkpq e^{i(\vecq+\vecG') \cdot \vecr} \rbkp}
         {(\eak - \eskq + \wwpid)(\ebkp - \etkpq + \wwpid)}
 \Bigr];\\
\Hx^{B-\textrm{res}}(\vecq;\omega) \equiv
 & - \frac{2}{\Omega} \sum_{abs\veck}\Bigl[ \
    \frac{ \lak e^{-i(\vecq+\vecG) \cdot \vecr} \rskq
        \lbk \ovxnl - \ovx \rak
        \lskq e^{i(\vecq+\vecG') \cdot \vecr} \rbk}
    {(\eak - \eskq + \wwpid)(\ebk - \eskq + \wwpid)}
 \Bigr]\\
 & + \frac{2}{\Omega} \sum_{ast\veck} \Bigl[ \
    \frac{ \lak e^{-i(\vecq+\vecG) \cdot \vecr} \rskq
        \lskq \ovxnl - \ovx \rtkq
        \ltkq e^{i(\vecq+\vecG') \cdot \vecr} \rak}
    {(\eak - \eskq + \wwpid)(\eak - \etkq + \wwpid)}
 \Bigr],
\end{split}
\end{equation}
\end{widetext}
where two is the spin factor, $\Omega$ is the crystal volume,
$\{a,b\}$ are valence bands, $\{s,t\}$ are conduction bands, the
matrix elements $\langle i;j |\owc| l;m \rangle$ are four-index
integrals defined as
\begin{equation}
\label{eq:4pt-integrals}
\begin{split}
 \langle i\veck+\vecq; j\veck' |\owc| l\veck'+\vecq; m\veck \rangle
    & \equiv \intsh d\vecr \! \intsh d\vecr' \,
    \phi_{i\veck+\vecq}^*(\vecr) \phi_{j\veck'}^*(\vecr')
    \\ & \times \wC(\vecr,\vecr') \phi_{l\veck'+\vecq}(\vecr) \phi_{m\veck}(\vecr'),
\end{split}
\end{equation}
$\ovxnl$ denotes an orbital-dependent exchange operator of the
form of the Hartree-Fock-like exchange operator but constructed
from the KS orbitals, i.e.,
\begin{equation}
\label{eq:HF-integrals}
 \langle i\veck+\vecq | \ovxnl | j\veck+\vecq \rangle
    \equiv - \sum_{a\veck'} \langle i\veck+\vecq; a\veck' |\owc| a\veck';
    j\veck+\vecq
    \rangle,
\end{equation}
and $\ovx$ is generated by the orbital-independent EXX KS
potential $v_x(\vecr)$. Here $\wC(\vecr,\vecr')$ yielding $\owc$
denotes the {\em generalized} Coulomb interaction, e.g.,
$1/|\vecr-\vecr'|$ for the bare Coulomb interaction.

Compared with the standard BSE approach, although we also have to
perform the four-index Coulomb integrals of Eqs.
(\ref{eq:4pt-integrals}) and (\ref{eq:HF-integrals}), we can
calculate optical spectra without diagonalizations in the space of
occupied and unoccupied single-particle states. Due to the huge
number of $\veck$-points involved in the optical spectrum
calculations, the size of the matrix to be diagonalized can be
very large, and the fact that we can avoid the diagonalization
process altogether in principle represents a significant numerical
advantage. In addition, note that $\Fx^{EXX}$ is free of Coulomb
singularities~\cite{FXEXX02} and thus the formal validity of the
EXX TDDFT approach does not depend on the consideration of the
thermodynamic limit. In the BSE equation, on the other hand,
integrable singularities appear which require special numerical
care~\cite{BSE98}.


We have implemented the {\em full dynamic} EXX kernel in
reciprocal space employing the plane wave basis. The accuracy of
the code has been carefully checked by numerically testing if the
calculated exchange kernel acts as a functional derivative of the
exchange potential with respect to the electron density at
$\omega=0$, $\dvx^{EXX} = \Fx^{EXX} \dn$. To obtain $\Fx^{EXX}$ we
first carried out self-consistent EXX ground-state calculations at
the experimental lattice constant of Si, $5.43 \AA$, and generated
the KS potential~\cite{Stadele97,DFT01}. Ten special-$\veck$
points and an orbital kinetic energy cutoff of 12.5 Ha have been
used. In the response calculation step, we solved the KS equations
once more at a larger number of $\veck$-points and obtained KS
orbitals and eigenvalues as the input for the construction of
$\chi_0$ and $\Fx^{EXX}$.  We adopted a shifted uniform
$\veck$-mesh, for which we employed up to $22 \times 22 \times 22$
$\veck$-points for $\chi_0$ and up to $9 \times 9 \times 9$
$\veck$-points for $\Fx^{EXX}$. The kinetic energy cutoff of 10 Ha
and $\delta$ = 0.15eV have been used, and ten conduction bands
were included.


The first question we addressed with small $\veck$-points sets
(e.g. $5 \times 5 \times 5$) was whether the adiabatic
approximation, defined as the $\omega = 0$ limit of Eqs.
(\ref{eq:Fx-EXX}) and (\ref{eq:Hx-resonant}), is justified.
Although the nature of the approximation within TDEXX is different
from that of the BSE approach in which the static screened
interaction is taken, we found that the adiabatic EXX kernel
generates an overall similar spectrum as the dynamic one.
Computing $\Fx^{EXX}$ only at a single $\omega$ results in a
significant reduction of computational workload, and thus the
converged calculation was performed in the adiabatic
approximation.

\begin{figure*}
\begin{minipage}[H]{0.70\linewidth}
  \centering\epsfig{file=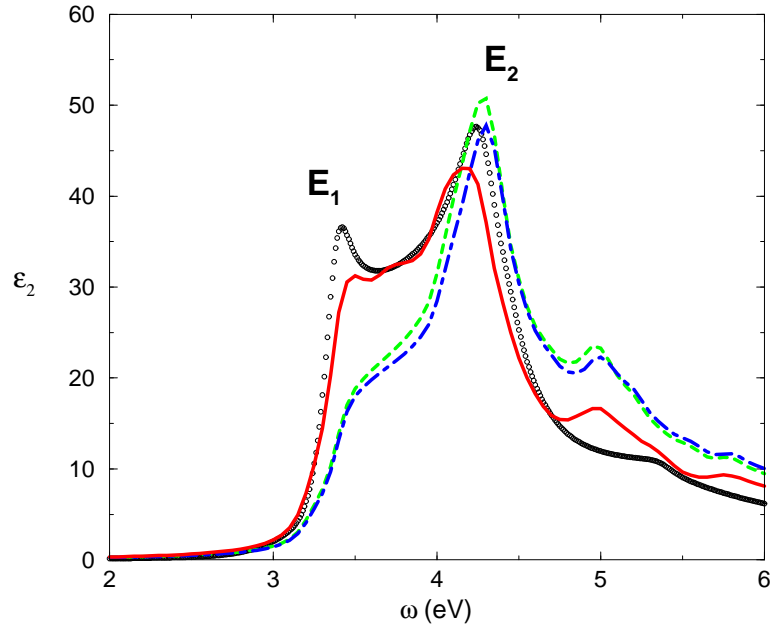,width=10cm}
\end{minipage}
\caption{\label{fig:TDEXX1} Calculated optical absorption spectrum
of Si from EXX (dashed lines), EXX+TDLDA (dot-dashed lines), and
EXX+TDEXX (solid lines). Circles denote experimental data of Ref.
\protect\onlinecite{Herzinger98}.}
\end{figure*}

Figure \ref{fig:TDEXX1} shows the absorption spectrum of Si
obtained at the single-particle EXX level and by taking into
account two-particle interaction effects via $\FHhat$ plus
$\Fxc^{LDA}$ (EXX+TDLDA) and via $\FHhat$ plus $\Fx^{EXX}$
(EXX+TDEXX). Compared with the much discussed LDA (and LDA+TDLDA)
spectrum, which is incorrectly shifted to the lower frequency
region by about 1 eV due to its well-known band gap
underestimation, the single-particle EXX absorption edge and
second ($E_2$) peak are in excellent agreement with the
experimental curve~\cite{Herzinger98} due to the realistic EXX
band structure~\cite{DFT01}. However, the first ($E_1$) peak
originating from electron-hole attractions is much underestimated
at the single-particle EXX level, and it is not recovered by
taking into account electron-hole interactions via TDLDA,
demonstrating the failure of $\Fxc^{LDA}$ to describe excitonic
effects. The EXX+TDEXX spectrum, on the other hand, is in
excellent agreement with the measured data: The absorption
strength at the $E_1$ peak region is correctly enhanced while that
of the higher frequency region is reduced, which shows that
$\Fx^{EXX}$ indeed provides the correct description of the
important excitonic effects.

To obtain the EXX+TDEXX spectrum in Fig. \ref{fig:TDEXX1}, we used
a slightly modified bare Coulomb interaction as $\wC$. Employing
the bare Coulomb interaction resulted in a collapse of the
spectrum due to too strong long-range Coulomb interaction of
electron-hole pairs at different $\veck$-points. We therefore cut
off the long-range Coulomb interaction for these pairs, or set to
zero the contributions to the integrals of Eq.
(\ref{eq:4pt-integrals}) for which $(\vecG + \veck - \veck')$ lies
within the first Brillouin zone~\cite{knekp}.  Devising a more
systematic strategy might be desirable in the future.

\begin{figure*}
\begin{minipage}[H]{0.70\linewidth}
  \centering\epsfig{file=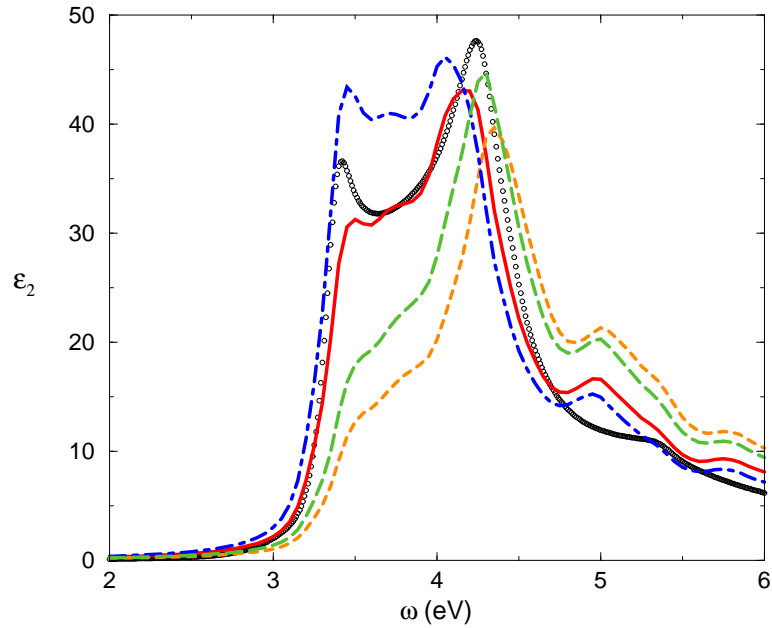,width=10cm}
\end{minipage}
\caption{\label{fig:TDEXX2} Calculated optical absorption spectrum
of Si from EXX+TDEXX with full EXX kernel (solid lines), $\Hx^B$
only (dashed lines), $\Hx^A$ only (dot-dashed lines), and full EXX
kernel with head and wings set to zero (long-dashed lines).
Circles: experiment.}
\end{figure*}

Now, to understand the underlying mechanism of the encouraging
EXX+TDEXX result, we analyze $\Fx^{EXX}$ by investigating the role
of different contributions to $\Hx$ as shown in Fig.
\ref{fig:TDEXX2}. First, by taking only $\Hx^B$, we obtain a
spectrum resembling that of the one-particle EXX but shifted
upward by about 0.3 eV.  No excitonic feature arises at this
level.  This spectrum is similar to that of the GWA, which is in
agreement with previous predictions~\cite{Gonze99,Tokatly01} that
GWA results should be recovered with the $\Hx^B$ part of the
screened EXX kernel. However, note that we have used an only
slightly modified bare Coulomb interaction, and, while the GWA
spectrum results from independent quasiparticles that involve ($N
\pm 1$)-electron excitations, the DFT ``quasiparticles" spectrum
already represents electron-number conserving ($N$-electron)
one-particle excitations at the time-independent
level~\cite{Gorling96,Umrigar98,DFT01}. Second, if we take only
$\Hx^A$, the spectrum is strongly enhanced at the lower frequency
region ($\lesssim$ 4 eV) while it is much reduced at the higher
frequency region ($\gtrsim$ 4 eV). The excitonic $E_1$ peak
appears predominantly while the $E_2$ peak is slightly red-shifted
by about 0.2 eV.  This shows that the Coulomb coupling of
electron-hole pairs is precisely the origin of the excitonic $E_1$
peak. Finally, we set the head and wings of $\Fx^{EXX}$ to zero
and thus make them $\vecq$-independent as in the case of the LDA
or GGA. By doing so, we obtain a spectrum similar to that of the
EXX+TDLDA without excitonic peaks or whatsoever.  This
demonstrates that taking into account the correct $\qzero$
behavior of $\Fxc$ is crucial.

We finally comment on other recent TDDFT works that have also
obtained optical spectra of semiconductors in good agreement with
experiments. First, de Boeij \textit{et al.} have left the domain
of conventional DFT and resorted to TD
current-DFT~\cite{deBoeij01}. Albeit they had to introduce an
empirical energy shift and a prefactor for the polarization
functional, their approach may represent a different route to
include the space nonlocality discussed above via a macroscopic
functional. However, our result shows that employing the current
response derived from a macroscopic functional is not necessary if
one properly incorporates the space nonlocality in the microscopic
density response. In this context, recent work of Reining
\textit{et al.} which was based on the idea of mapping the BSE to
the TDDFT is in agreement with ours~\cite{Reining02}. But in their
scheme they first invoked the GWA quasiparticle shift and second
employed an empirical static scaled Coulomb kernel as $\Fxc$.
Although their approach is not rigorous, we note that their simple
static kernel nevertheless exhibits a correct $\qzero$ divergence.


In summary, we have reported calculations of the optical
absorption spectrum of Si with the novel TDEXX approach. Of
particular interest was the nature of the dynamic and nonlocal
orbital-based EXX kernel which exhibits the $\qzero$ behavior of
the exact $\Fxc$ and successfully generated excitonic features in
the spectrum. We showed that including Coulomb coupling of KS
electron-hole pairs at different $\veck$-points and the correct
long-wavelength behavior of the \xc\ kernel is crucial, while the
adiabatic approximation can be justified.

We thank L. Reining, L. X. Benedict, and E. L. Shirley for helpful
discussions. This work has been supported by the Humboldt
Foundation (Y.-H.K.), the Deutsche Forschungsgemeinschaft, and the
Fonds der Chemischen Industrie (A.G.).


\newpage

\begin{thebibliography}{}
\bibitem{AU}
We use Hartree atomic units, $\hbar = e = m_e = 4\pi \epsilon_0 =
1$, throughout.
\bibitem{convention}
$\Hx^A$ and $\Hx^B$ represents ($\Hx^1+\Hx^2$) and ($\Hx^3+\Hx^4$)
of Ref. [\onlinecite{FXEXX02}]. We assume that the external
perturbation and other quantities have the time dependence of
$e^{-i \omega t} e^{\delta}$, where $\delta \rightarrow 0^+$ is a
convergence factor.
\bibitem{knekp}
This neglect of long-range contribution is necessary only for
$\veck \neq \veck'$, because $\veck = \veck'$ contributions with
$|\vecG + \veck - \veck'| = |\vecG |= 0$ which would lead to
Coulomb singularities do not occur in the EXX kernel expression.
\end{thebibliography}
\end{document}